\documentclass[letter, longauth, traditabstract]{aa} 
\usepackage{natbib}
\usepackage{graphicx}
\usepackage{txfonts}
\usepackage{threeparttable}
\def\Msun{\hbox{$M_{\odot}$}}               
\def\Lsun{\hbox{$L_{\odot}$}}               
\def\Rstar{\hbox{$R_{\star}$}}              
\def\Mdot{\hbox{$\dot{M}$}}               
\def\Teff{\hbox{$\rm{T}_{\rm eff}$}}            
\def\arcsec{\hbox{$^{\prime\prime}$}}

\begin{document}
  \title{PACS and SPIRE Spectroscopy of the Red Supergiant VY CMa
\thanks{Herschel is an ESA space observatory with science instruments provided by European-led Principal Investigator consortia and with important participation from NASA. 
}}

  \author{
         P. Royer\inst{1}
         \and
         L. Decin\inst{1,12}
         \and
         R. Wesson\inst{2}
         \and
         M.J. Barlow\inst{2}
         \and
         E.T. Polehampton\inst{7,9}
         \and
         M. Matsuura\inst{2,15}
         \and
         M. Ag\'undez\inst{14,16}
         \and
         J.A.D.L. Blommaert\inst{1}
         \and
         J. Cernicharo\inst{14}
         \and
         M. Cohen\inst{4}
         \and
         F. Daniel\inst{14}
         \and
         P. Degroote\inst{1}
         \and
         W. De Meester\inst{1}
         \and
         K. Exter\inst{1}
         \and
         H. Feuchtgruber\inst{3}
         \and
         W.K. Gear\inst{5}
         \and
         H.L. Gomez\inst{5}
         \and
         M.A.T. Groenewegen\inst{11}
         \and
         P.C. Hargrave\inst{5}
         \and
         R. Huygen\inst{1}
         \and
         P. Imhof\inst{10}
         \and
         R.J. Ivison\inst{6}
         \and
         C. Jean\inst{1}
	\and
	F. Kerschbaum\inst{13}
         \and
         S.J. Leeks\inst{7}
         \and
         T. Lim\inst{7}
         \and
         R. Lombaert\inst{1}
         \and
         G. Olofsson\inst{8}
	\and
	T. Posch\inst{13}
         \and
         S. Regibo\inst{1}
         \and
         G. Savini\inst{2}
         \and
         B. Sibthorpe\inst{6}
         \and
         B.M. Swinyard\inst{7}
         \and
         B. Vandenbussche\inst{1}
         \and
         C. Waelkens\inst{1}
         \and
         D.K. Witherick\inst{2}
         \and
         J.A. Yates\inst{2}
         }


  \institute{Instituut voor Sterrenkunde,
             Katholieke Universiteit Leuven, Celestijnenlaan 200D, 3001 Leuven, Belgium\\
	      \email{pierre@ster.kuleuven.be}
       \and
             Dept of Physics \& Astronomy, University College London, Gower St, London WC1E 6BT, UK\\
       \and
             Max-Planck-Institut f\"ur extraterrestrische Physik, Giessenbachstrasse, 85748, Germany\\
       \and
             Radio Astronomy Laboratory, University of California at Berkeley, CA 94720, USA\\
       \and
             School of Physics and Astronomy, Cardiff University, 5 The Parade, Cardiff, Wales CF24 3YB, UK\\
       \and
             UK Astronomy Technology Centre, Royal Observatory Edinburgh, Blackford Hill, Edinburgh EH9 3HJ, UK\\
       \and
             Space Science and Technology Department, Rutherford Appleton Laboratory, Oxfordshire, OX11 0QX, UK\\
       \and
             Dept of Astronomy, Stockholm University, AlbaNova University Center, Roslagstullsbacken 21, 10691 Stockholm, Sweden\\
       \and
            Department of Physics, University of Lethbridge, Lethbridge, Alberta, T1J 1B1, Canada\\
       \and
            Blue Sky Spectroscopy, 9/740 4 Ave S, Lethbridge, Alberta T1J 0N9, Canada \\
       \and
            Royal Observatory of Belgium, Ringlaan 3, B-1180 Brussels, Belgium  \\
 		\and
	Sterrenkundig Instituut Anton Pannekoek, University of Amsterdam, Science Park 904, NL-1098 Amsterdam, The Netherlands \\
          \and
    	    University of Vienna, Department of Astronomy, T{\"u}rkenschanzstra\ss{}e 17, A-1180 Vienna, Austria\\
          \and
            Astrophysics Dept, CAB (INTA-CSIC), Crta Ajalvir km4, 28805 Torrejon de Ardoz, Madrid, Spain\\
          \and
            Mullard Space Science Laboratory, University College London, 
Holmbury St. Mary, Dorking, Surrey RH5 6NT, United Kingdom \\    
          \and
            LUTH, Observatoire de Paris-Meudon, LERMA UMR CNRS 8112, 5
            place Jules Janssen, 92195 Meudon Cedex, France }

  \date{Received 01 April 2010 - Accepted 20 April 2010 }

\abstract{With a luminosity $>10^5$ \Lsun\ and a mass-loss rate of $\sim 2. 10^{-4}$ \Msun\,yr$^{-1}$, the red supergiant VY CMa truly is a spectacular object. Because of its extreme evolutionary state, it could explode as supernova any time. Studying its circumstellar material, into which the supernova blast will run, provides interesting constraints on supernova explosions and on the rich chemistry taking place in such complex circumstellar envelopes. We have obtained spectroscopy of VY\,CMa over the full wavelength range offered by the PACS and SPIRE instruments of Herschel, i.e. 55 -- 672 micron. The observations show the spectral fingerprints of more than 900 spectral lines, of which more than half belong to water. In total, we have identified 13 different molecules and some of their isotopologues. A first analysis shows that water is abundantly present, with an ortho-to-para ratio as low as $\sim$1.3:1, and that chemical non-equilibrium processes determine the abundance fractions in the inner envelope.}{}{}{}{}

  \maketitle
%

\section{Introduction}

The oxygen-rich red supergiant VY\,CMa is one of the most extreme objects of its kind. Its large mass-loss rate results in a very dense circumstellar envelope and an enormous infrared excess \citep[][]{2001ApJ...557..844H}. Infrared observations are invaluable for in-depth studies of the complex chemical and dynamical processes at work in its circumstellar envelope. By this means, VY\,CMa has been shown to harbour a rich, water-dominated chemistry \citep[e.g.][]{1999ApJ...517L.147N, Polehampton2010A&A...510A..80P}. The unprecedented wavelength coverage and spectral resolution of the Herschel instruments will undoubtedly push our understanding of the mass-loss related phenomena a step further.

In this context, VY\,CMa has been observed as part of the Herschel   Guaranteed Time Key Programme ``Mass-loss of Evolved StarS'', MESS (Groenewegen et al., {\it in prep.}). The spectra discussed hereafter have been obtained with the PACS and SPIRE instruments of Herschel \citep{Pilbratt2010}. The instruments and their in-orbit performance are described in \citet{Poglitsch2010} and \citet{Griffin2010}. The SPIRE calibration methods and accuracy are outlined by \citet{Swinyard2010}.

\begin{figure*}
\centering
\includegraphics[width=\textwidth,height=16.3cm]{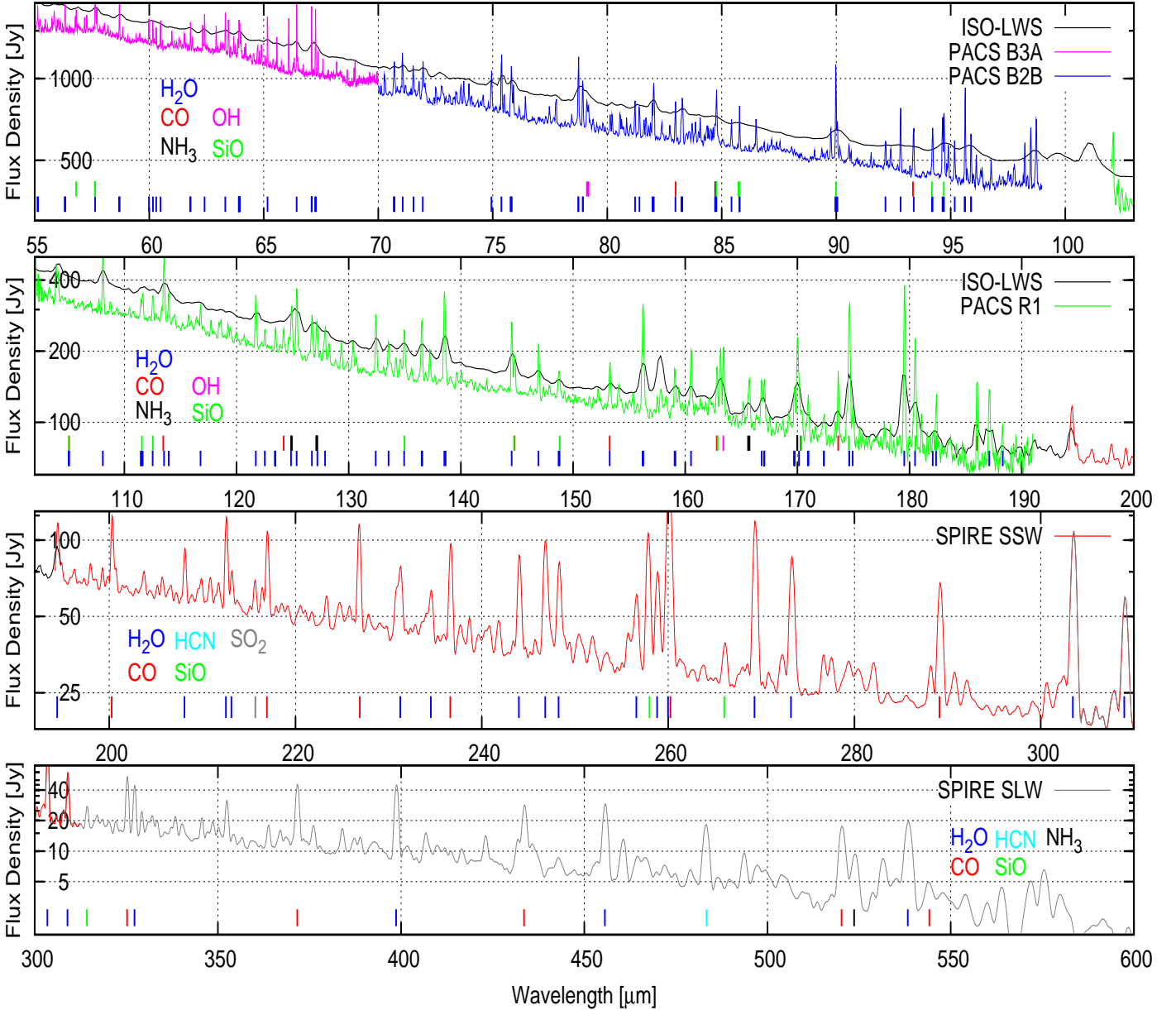}
\caption{The complete PACS + SPIRE spectrum of VY~CMa. The various spectral bands are presented in order of increasing wavelength. For the sake of clarity, PACS band B2A is not included, as it matches the wavelength range of PACS-B3A. The ISO-LWS spectrum is overplotted for comparison \citep{Polehampton2010A&A...510A..80P}. For PACS, the spectrum of the central spaxel is displayed, after application of the point source correction (version 2.0). The strongest spectral features of the main molecules are indicated.}
              \label{FigFullSpectrum}
\end{figure*}

\section{Observations and data reduction}
The PACS observations of VY\,CMa consist of 5 spectral energy distribution (SED) scans, acquired on 3 and 13 November 2009 (see Table~\ref{tab_obs} for details). All observations were performed with a non-standard version of the PACS-SED observing mode. The difference with the standard mode resides in the wavelength coverage, as our SEDs cover the complete PACS wavelength range in the four bands, in two observations. All observations were ``chop-nodded'', and contained one single nodding cycle, and one single up-down scan in wavelength.

The data were reduced with the nominal pipeline, except that we combined the spectra obtained at the two nod positions after rebinning only. The flux calibration is based on a constant detector response and the wavelength ranges affected by light leaks have been excluded from the present analysis.

The original design of the PACS SED observing mode included Nyquist spectral sampling. However, in the early version that was used for these observations, a few wavelength ranges were still slightly undersampled. Consequently, rebinning the spectra at the nominal resolution of the spectrometer returns a number of empty bins. For the modeling, we worked at the nominal resolution, and used linear interpolation at the wavelengths of the empty bins. For the line identifications, we have rebinned the spectrum to a slightly degraded resolution (half Nyquist)\footnote{The standard PACS-SED was modified on OD\,305, and the full Nyquist sampling has been implemented since then.}.

\begin{table*}[hbt]
\caption[Journal]{Journal of observations. The target coordinates are $\alpha_{2000}$:\,07h22m58.3s, $\delta_{2000}$:\,-25$^\circ$46'03.2''. On Operational Day 183, the PACS observations were 2$\times$2 rasters with a step of 4\arcsec\ along each of the instrument axes (Fig.~\ref{fig_obs}).}
\label{tab_obs}
\centering
\begin{tabular}{@{} l c c l c c c l l @{}}
\hline
\textbf{Instr.} & \textbf{OBSID}  &  \textbf{OD} & \textbf{Date} & \textbf{Pointing} &\textbf{Position Angle} & \textbf{Duration} & \multicolumn{2}{c}{\textbf{Wavelengths}}\\
                &                 &              &               & \textbf{Mode}     &    [Degrees]           &  [Seconds]        & \multicolumn{2}{c}{[$\mu$m]}   \\
\hline
PACS & 1\,342\,186\,653  & 173 & 03 Nov 2009 & Single & 112 &  3655 & [54-73] & [102-220] \\
PACS & 1\,342\,186\,654  & 173 & 03 Nov 2009 & Single & 112 &  2759 & [67-109.5] & [134-219] \\
PACS & 1\,342\,186\,994  & 183 & 13 Nov 2009 & Raster & 120 & 14745 & [54-73]& [102-220] \\
PACS & 1\,342\,186\,995  & 183 & 13 Nov 2009 & Raster & 120 & 11167 & [67-109.5]& [134-219] \\
SPIRE & 1\,342\,183\,813 & 123 & 13 Sep 2009 & Single &  73 &  4249 & [194-313]& [303-671] \\
\hline
\end{tabular}
\end{table*}

The SPIRE spectrum of VY~CMa was obtained on Operational Day (OD)\,123 and consisted of 30 repetitions. Each repetition consisted of 1 forward and 1 reverse scan of the FTS. The total on-source integration time was 3996\,sec. The unapodized spectral resolution was 1.4\,GHz (0.048\,cm$^{-1}$), after apodization \citep[using extended Norton-Beer function 1.5;][]{Naylor2007JOSAA..24.3644N} this is 2.1\,GHz (0.07\,cm$^{-1}$). 

\begin{figure}
\centering
\includegraphics[width=9cm]{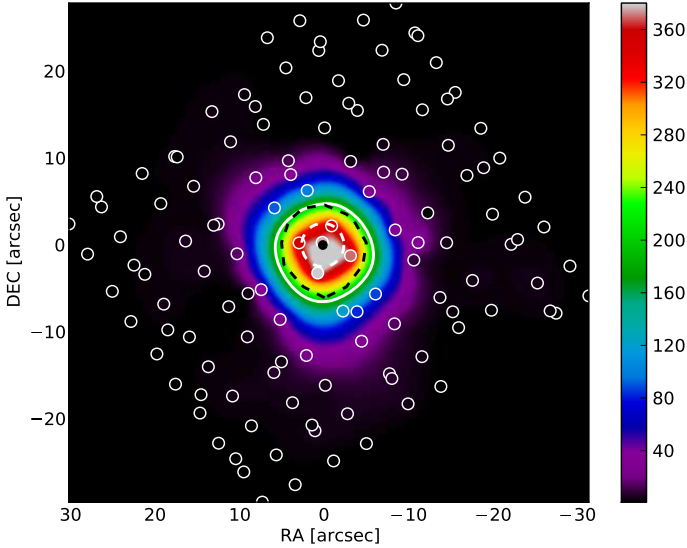}
\caption{Line peak intensity, expressed in Jy, for the o-H$_2$O\,($7_{1,6}$-$6_{2,5}$) line at 66.09\,$\mu$m, in all PACS observations. The dashed contours represent in white, the diffraction limited telescope beam at 66\,$\mu$m (5\arcsec\ diameter), and in black, the instrumental PSF at half maximum. The PSF is significantly wider due to the spatial undersampling of the beam (the spaxels are 9.4\arcsec$\times$9.4\arcsec). The outer white contour is at 50\% of maximum H$_2$O intensity. The emission is not resolved. The shape visible at low flux corresponds to the PSF wings. The white circles indicate the centers of the spaxels during each of the SED scans. The black dot in the center marks the object coordinates.}
\label{fig_obs}
\end{figure}

\section{Results}
\subsection{Observational results}
Figure\,\ref{FigFullSpectrum} shows the full PACS and SPIRE spectrum of VY\,CMa. More than 930 molecular emission lines have been detected, 95\% of which we could already identify. More than 400 of these lines originate from water vapour, but many different molecules contribute to the spectra: the identified species and their isotopologues are $^{12}$CO, $^{13}$CO, C$^{17}$O,  C$^{18}$O,  ortho-H$_2^{16}$O (o-H$_2^{16}$O), para-H$_2^{16}$O (p-H$_2^{16}$O),  o-H$_2^{17}$O, o-H$_2^{18}$O, p-H$_2^{18}$O, CN, CS, SO, NH$_3$, OH, $^{28}$SiO, $^{29}$SiO, $^{30}$SiO, SiS,  HCN, SO$_2$,  H$_2$S, [C{\sc ii}]. H$_3$O$^+$ is possibly present as well, as reported by \citet{Phillips1992ApJ...399..533P} and \citet{Polehampton2010A&A...510A..80P}, but due to the high density of lines, detailed modeling is needed to give a definitive answer (Matsuura et al. 2010, {\it in prep}). Both low-excitation (e.g., CO J=4-3, at 55\,K) and high-excitation (e.g., o-H$_2$O\,($16_{8,9}$-$15_{9,6}$) at 5500\,K) are found, allowing us to study the molecular envelope from the outermost layers down to the dust formation region close to the star. In this way, the thermodynamical structure of the complex envelope and the chemical processes in the envelope can be determined.

The flux calibrations of the two instruments match within 15\%, the different spectral ranges of a given instrument agree within 5\%. The fluxes measured by PACS and IRAS at 60 and 100\,$\mu$m are similar within 5 and\ 15\%.  The agreement between SPIRE and ISO-LWS is within a few percent at 195$\mu$m, and the PACS and ISO-LWS spectra coincide to 20\,$\pm$\,10\% over all wavelengths. Note that the ISO-LWS fluxes were found to be high in comparison with IRAS. \citet{Gry2003sws..book.....G} reported that the LWS fluxes were 20\% above those of IRAS at the flux level of VY\,CMa. On the other hand, the PACS flux in Fig.\ref{FigFullSpectrum} might be slightly underestimated, due to the extension of the source.

Indeed, although it appears unresolved at first sight, in lines as well as in continuum (Fig.~\ref{fig_obs}), VY\,CMa can not be considered as a point source for PACS. Its flux, integrated over the field, shows an excess of 15$\pm$3\% in the blue and 10$\pm$5\% in the red compared to the flux measured in the central spatial pixel (spaxel) after application of the point source correction, for the observations of OD\,173 (shown in Fig.~\ref{FigFullSpectrum}). The same comparison, carried out on the Neptune SED that was obtained on the same OD shows Neptune ($\sim$2\arcsec\ diameter during the observation) to be compatible with a point source, hence confirming the extension of the emission around VY~CMa. The ratio between central spaxel and integrated flux was computed for all raster positions. All of these ratios show comparable or larger excesses, hence ruling out the possibility that, on OD\,173, the satellite was pointing significantly out of the peak of emission.

Additionally, these central to integrated flux ratios, as well as dedicated measurements in the spectra, reveal that the line to continuum ratios are constant to within 5\% over the whole field of view, at all PACS wavelengths, the only exception being the [C{\sc ii}] line at 157.74\,$\mu$m, which has a stronger integrated flux due to its interstellar origin. Finally, we created maps of the line ratios between high and low-excitation transitions of water, e.g.\  o-H$_2$O\,($7_{1,6}$-$6_{2,5}$) 66.09$\mu$m / o-H$_2$O\,($3_{3,0}$-$2_{2,1}$) 66.43$\mu$m in the blue and o-H$_2$O\,($7_{3,4}$-$7_{2,5}$) 166.82$\mu$m / o-H$_2$O\,($2_{1,2}$-$1_{0,1}$) 179.53$\mu$m in the red part of the PACS SED.
These maps show no structure either, supporting the interpretation that all the emission is localized in the central part of our field. Hence, although detailed comparison with a high quality PSF is still to be performed,  all PACS observations are consistent with an extension of 5 to 10\arcsec, which is compatible with the central part of the nebula as imaged in the visible, near and mid-infrared by \citet{Smith2001AJ....121.1111S}.

\subsection{Modeling the envelope: CO, H$_2$O, HCN and SiO}

\begin{figure*}[hbtp]
\centering
\includegraphics[width=\textwidth, height=5cm]{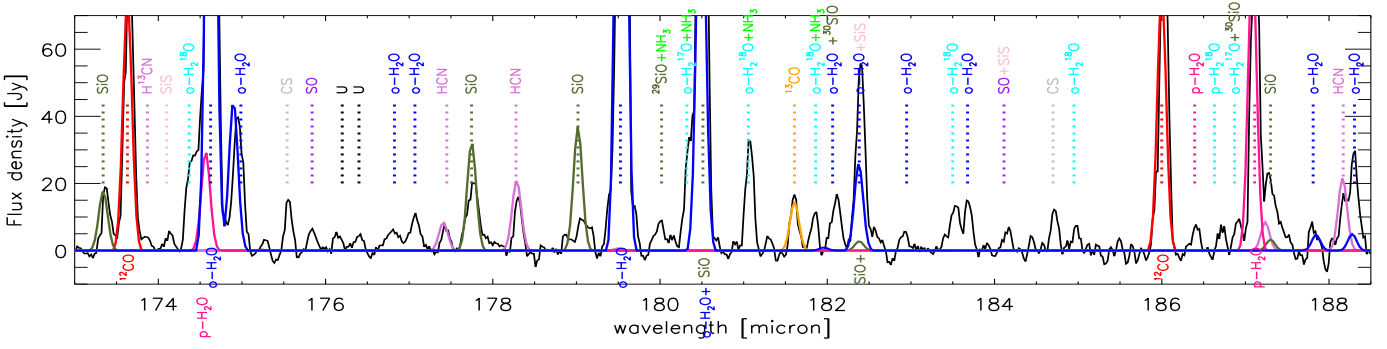}
\caption{The continuum-subtracted PACS spectrum (black) between 173 and 188.5\,$\mu$m. The main contributing molecules and isotopes are identified. Features not yet identified are indicated with a 'U'.  The first modeling results are shown, with one color per molecular species.}
\label{fig_lineIds}
\end{figure*}

Due to the rich water spectrum, most of the emission lines are blended. In addition, many line transitions have high optical depth. Consequently, the fractional molecular abundance can only be retrieved from a proper solution of the radiative transfer equation coupled to the statistical equilibrium equations. Although we realize that the envelope around VY~CMa has a very complex structure \citep{Smith2009AJ....137.3558S}, we have assumed a spherically symmetric envelope. Using this assumption, the average density over all directions of $^{12}$CO, $^{13}$CO, HCN, o-H$_2^{16}$O, p-H$_2^{16}$O, and $^{28}$SiO can be determined. A more detailed study will be presented in Matsuura et al.\ (\emph{in prep.}).

We have used the non-local thermodynamic equilibrium (non-LTE) radiative transfer code GASTRoNOoM \citep{Decin2006A&A...456..549D, Decin2010}, with  line lists and collisional rates as specified by \citet{Decin2010}. The kinetic temperature is calculated from the balance of cooling and heating, with the main contributions coming from adiabatic expansion, grain-gas collisions, and CO and H$_2$O ro-vibrational transitions. The conservation of momentum delivers the velocity structure. The model parameters are summarized in Table~\ref{Table:1},  a comparison between observed data and theoretical predictions is shown in Fig.~\ref{fig_lineIds}. Specifically, we derive a mean mass-loss rate of $1.5 \times 10^{-4}$\,\Msun\,yr$^{-1}$, a fractional abundance of SiO of $3 \times 10^{-5}$, of HCN we find $3\times10^{-6}$ and of o-H$_2$O being $3\times10^{-4}$. We derive a $^{12}$CO/$^{13}$CO isotopic ratio of 60, and an ortho-to-para water ratio (OPR) of 1.27:1. Due to the high water optical depth, the latter remains rather uncertain.

\begin{table}
 \caption{Model parameters for VY~CMa. Numbers in italics are input parameters that have been kept fixed at the given value.}
\label{Table:1}
\begin{center}
\vspace*{-1.5ex}
\begin{tabular}{lc|lc}
\hline \hline
\rule[0mm]{0mm}{5mm}\Teff [K] & \emph{2800}$^{a}$ & $R_{\rm{dust}}$ [\Rstar] & \emph{10}$^a$ \\
$R_{\star}$ [$10^{14}$\,cm] & \emph{1.6}$^{a}$ &  $^{12}$CO/$^{13}$CO & 60 \\
$[$CO/H$_2]$ [$10^{-4}$]    & \emph{3} & n(o-H$_2$O)/n(H$_{\rm{tot}}$) & $3 \times 10^{-4}$  \\
distance  [pc] & \emph{1140}$^b$ &  n(SiO)/n(H$_{\rm{tot}}$) & $3 \times 10^{-5}$  \\ 
$v_{\infty}$     [km\,s$^{-1}$] &  \emph{35} $^{a}$&  n(HCN)/n(H$_{\rm{tot}}$) & $3 \times 10^{-6}$\\
\Mdot$(r)$ [\Msun\,yr$^{-1}$] & $1.5 \times 10^{-4}$ & (o-H$_2$O)/(p-H$_2$O) & 1.27:1\\
\hline
\end{tabular}
\end{center}
\vspace{-2ex}
$^{a}$\citet{Decin2006A&A...456..549D}, $^{b}$\citet{Choi2008PASJ...60.1007C} 
\end{table}
 
\section{Discussion}

{\bf HCN:} \citet{Ziurys2009ApJ...695.1604Z} derived a HCN fractional abundance [HCN/H$_2$] of $1.2 \times 10^{-6}$ in the spherical outflow and of $7.5 \times 10^{-6}$ in the so-called red and blue flow for a mass-loss rate of $2\times10^{-4}$\,\Msun\,yr$^{-1}$ and a distance of 1500\,pc, being similar to our results. This abundance is orders of magnitude higher than the thermodynamic equilibrium value (TE) [HCN/H$_{\rm{tot}}$] of $6 \times 10^{-11}$ \citep{Duari1999A&A...341L..47D}. A similar high abundance of HCN has also been derived for oxygen-rich (O-rich) asymptotic giant branch (AGB) stars \citep{Ziurys2009ApJ...695.1604Z, Decin2010}. Photochemical predictions for the outer envelope of O-rich AGB stars \citep{Willacy1997A&A...324..237W} are clearly too low by at least an order of magnitude. Pulsation driven shock models can account for this high concentration of HCN, yielding a HCN fractional abundance in the inner wind (between 2.2 and 5\,\Rstar) ranging from $2 \times 10^{-6}$ to $1 \times 10^{-5}$ \citep{Duari1999A&A...341L..47D, Cherchneff2006A&A...456.1001C}.

{\bf SiO:} The derived SiO abundance is consistent with the results of \citet{Ziurys2007Natur.447.1094Z}. Both inner wind shock-induced non-equilibrium chemistry \citep{Duari1999A&A...341L..47D, Cherchneff2006A&A...456.1001C} and outer wind photochemical processes \citep{Willacy1997A&A...324..237W}  predict almost the same abundance value.

{\bf H$_2$O:} \citet{Zubko2004ApJ...610..427Z} derived a water abundance [H$_2$O/H$_2$] of $4 \times 10^{-4}$ from fitting ISO SWS and LWS observations, with an OPR of 1:1, similar to our results. Low OPR ratios are also derived for the hypergiant NML Cyg by \citet{Zubko2004ApJ...610..427Z}  and for the O-rich AGB star W Hya (1:1.3 by \citet{Zubko2000ApJ...544L.137Z} and 1:1 by \citet{Barlow1996A&A...315L.241B}). An OPR close to 1:1 reflects a cold spin temperature, around 15\,K \citep{Mumma1987A&A...187..419M}. It is thought that the OPR value should correspond to the physical conditions where water was formed, i.e. the warm and dense regions of the envelope where the chemistry is under thermodynamical equilibrium and controlled by three body reactions. Hence, we could expect an OPR value of 3. The large opacity of the water lines very sensibly reduces our sensitivity to the OPR value, as also noted by \citet{Barlow1996A&A...315L.241B}, so that, even using H$_2^{18}$O, we can at present not rule out the aforementioned expected value.

{\bf $^{12}$CO/$^{13}$CO:} From low-excitation CO lines, \citet{Muller2007ApJ...656.1109M} derived also a $^{12}$CO/$^{13}$CO ratio of 60, while \citet{Milam2009ApJ...690..837M} obtained a $^{12}$C/$^{13}$C value of 46 for the spherical wind. This indicates that the isotopologue ratio is (roughly) constant in the spherical wind.

\section{Conclusion}

Through the detection of more than 900 emission lines of various chemical species in VY\,CMa, PACS and SPIRE have shown their diagnostic strength to unravel the chemically complex mass-loss processes at work in evolved stars. Specifically, we have derived the HCN, SiO and water fractional abundances of the circumstellar material around VY\,CMa, which also shows a suprisingly low ortho-to-para water ratio, close to 1.  The origin of the latter is not yet understood, but the Herschel observations of many evolved stars in the MESS programme might shed light on this topic.

\begin{acknowledgements}
PACS has been developed by a consortium of institutes led by MPE (Germany) and including UVIE (Austria); KUL, CSL, IMEC (Belgium); CEA, OAMP (France); MPIA (Germany); IFSI, OAP/AOT, OAA/CAISMI, LENS, SISSA (Italy); IAC (Spain). This development has been supported by the funding agencies BMVIT (Austria), ESA-PRODEX (Belgium), CEA/CNES (France), DLR (Germany), ASI (Italy), and CICT/MCT (Spain). SPIRE has been developed by a consortium of institutes led by Cardiff Univ. (UK) and including Univ. Lethbridge (Canada); NAOC (China); CEA, LAM (France); IFSI, Univ. Padua (Italy); IAC (Spain); Stockholm Observatory (Sweden); Imperial College London, RAL, UCL-MSSL, UKATC, Univ. Sussex (UK); Caltech, JPL, NHSC, Univ. Colorado (USA). This development has been supported by national funding agencies: CSA (Canada); NAOC (China); CEA, CNES, CNRS (France); ASI (Italy); MCINN (Spain); SNSB (Sweden); STFC (UK); and NASA (USA). MG, JB, WDM, KE, RH, CJ, SR, PR \& BV acknowledge support from the Belgian Federal Science Policy Office via the PRODEX Programme of ESA. LD  acknowledges support from the Fund for Scientific Research - Flanders (FWO). 
FK acknowledges funding by the Austrian Science Fund FWF under project numbers P18939-N16 \& I163-N16.
The authors thank D.\ Ladjal \& B.\ Acke for their support.
\end{acknowledgements}

\bibliographystyle{aa}
\bibliography{14641}

\end{document}